\documentclass[journal, ,onecolumn,12pt]{IEEEtran}
\usepackage{graphicx} 
\usepackage{xcolor}
\usepackage{amsmath} 
\usepackage{amsfonts}
\begin{document}

\title{Conformal Calibration: Ensuring the Reliability of Black-Box AI  in Wireless Systems}

\author{Osvaldo Simeone, Sangwoo Park, and Matteo Zecchin
\vspace{-0.5cm} 
\thanks{The authors are with the King’s Communications, Learning and Information Processing (KCLIP) lab within  the Centre for Intelligent Information Processing Systems (CIIPS) at the Department of Engineering,  King’s College London, London, WC2R 2LS, UK. (email: \{osvaldo.simeone, sangwoo.park, matteo.1.zecchin\}@kcl.ac.uk). This work was supported by the European Union’s Horizon Europe project CENTRIC (101096379),  by~an Open Fellowship of the EPSRC (EP/W024101/1), and~by the EPSRC project (EP/X011852/1).}
}

\maketitle

\begin{abstract}

AI is poised to revolutionize telecommunication networks by boosting efficiency, automation, and decision-making. However, the black-box nature of most AI models introduces substantial risk, possibly  deterring adoption by network operators. These risks are not addressed by the current prevailing deployment strategy, which typically follows a best-effort train-and-deploy paradigm.  This paper reviews conformal calibration, a general framework  that moves beyond  the state of the art  by adopting computationally lightweight, advanced statistical tools that offer {formal reliability guarantees} without requiring further training or fine-tuning.  Conformal calibration encompasses pre-deployment calibration via uncertainty quantification or hyperparameter selection; online monitoring to detect and mitigate failures in real time; and counterfactual post-deployment performance analysis to address ``what if'' diagnostic questions after deployment.  By weaving conformal calibration into the AI model lifecycle, network operators can establish confidence in black-box AI models as a dependable enabling technology for wireless systems.
\end{abstract}



\section{Introduction}

\subsection{Motivation}

Next-generation wireless networks are expected to leverage AI for tasks ranging from physical-layer processing to resource management. Initiatives like O-RAN exemplify this trend by defining open network architectures that enable data-driven control  at different time scales  via modular AI applications \cite{bonati2021intelligence}. While AI promises improved efficiency and flexibility, most AI apps function as \emph{black boxes},  raising significant reliability concerns.  These reliability concerns may make operators hesitant to cede network functionalities to black-box systems without additional safeguards.





As illustrated in Figure 1, modern wireless network architectures -- such as O-RAN -- implement key network functionalities through AI applications executed on controllers that interface with underlying network elements. These AI apps are often developed by diverse third-party vendors, while their lifecycle -- both pre- and post-deployment -- is managed by higher-level controllers within the network. The prevailing deployment strategy typically follows a best-effort train-and-deploy paradigm: the higher-level controllers select apps for deployment based on the assumption that the training data used reflects traffic and connectivity conditions similar to those currently observed in the network.


This paper reviews, and advocates for, solutions that move beyond  best-effort state-of-the-art strategies by adopting advanced statistical tools for the \emph{calibration} of black-box models \cite{angelopoulos2024theoretical,ramdas2024hypothesis}. We refer to these methods broadly as \emph{conformal calibration}. The general goal of conformal calibration is that of ensuring \emph{formal reliability guarantees} without requiring further training or fine-tuning.  By deploying  conformal calibration methods around existing AI apps, network operators can ensure that \emph{key performance indicators} (KPIs) are satisfied with a user-defined confidence level, which will be denoted by a network-specified probability $1-\alpha$. 

For example, the network operator may wish the latency experienced by  a class of devices to be below $1$ ms for an average fraction of at least  $1-\alpha=0.9999$ users; or it may require the rate provided to a virtual reality service to be above 1 Gbit/s for a fraction of at least  $1-\alpha=0.9$ of the time.


\begin{figure}
    \centering
    \includegraphics[width=0.6\textwidth]{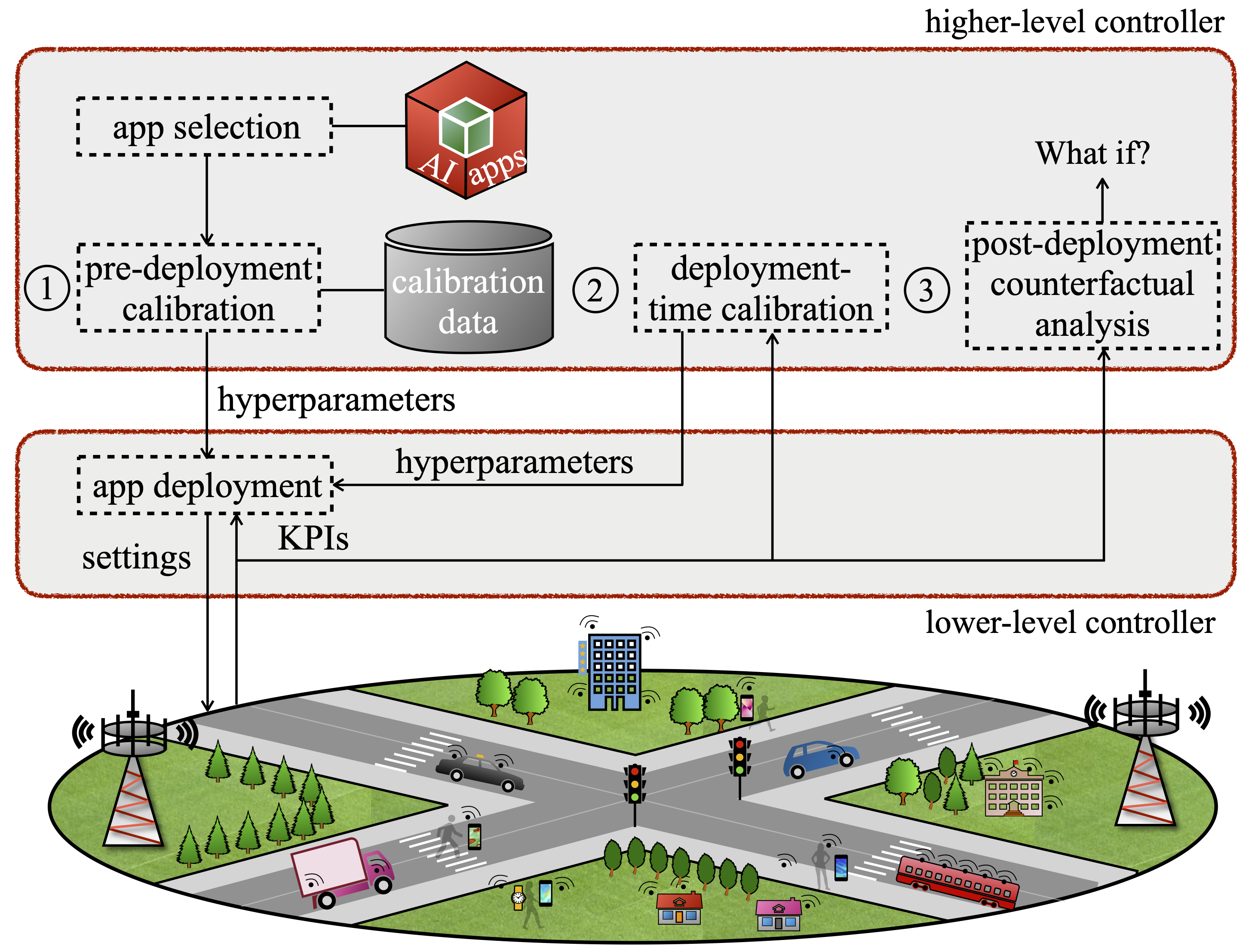}
    \caption{This paper describes conformal calibration, a general framework aimed at ensuring the reliability of black-box AI models for wireless systems. While the AI models are run at controllers that are directly connected to the network elements, conformal calibration is implemented at a higher-level controller. The framework supports the full lifecycle of AI applications, encompassing three distinct phases: pre-deployment calibration, deployment-time calibration, and post-deployment counterfactual analysis.  }
    \label{fig:setup}
\end{figure}

\subsection{Calibration and the Lifecycle of an AI App}

 Like the prevailing train-and-deploy paradigm, conformal calibration does not require altering the pre-trained AI models available for deployment at the higher-level controllers. Instead, it leverages the controller’s ability to set \emph{hyperparameters} through the AI application’s API and/or \emph{post-process} the model outputs, thus integrating seamlessly with existing deployment pipelines.

  Examples of hyperparameters include \emph{architectural} choices, such as the selection of pre-designed submodules; \emph{inference} parameters, such as fairness weights for scheduling and the temperature used to control the level of randomness in decision-making or data generation;  and \emph{deployment} parameters, such as arithmetic precision \cite{mungari2024oreo}.   Note that our focus is on test-time hyperparameters -- those adjustable at inference or deployment -- rather than training-time hyperparameters like learning rates or training schedules, though the statistical techniques discussed here are applicable to both.

Within the conformal calibration framework, post-processing  involves lightweight, inexpensive  operations, such as computing KPI statistics and applying thresholding nonlinearities to enable uncertainty quantification.  
  
 As illustrated in Figure 1, in this paper,  we review three key phases of the calibration process aiming ensuring  reliability across the entire \emph{lifecycle} of an AI model in wireless systems.

 \noindent $\bullet$ \emph{Pre-deployment calibration}: Prior to the deployment of an AI app, a higher-level  controller  determines suitable post-processing strategies and/or hyperparameter settings so as to ensure \emph{statistical KPI guarantees} of the form $\Pr[\textrm{KPI target violated}]\leq \alpha$. Accordingly, the operator is ensured that, under \emph{nominal} conditions, the probability of violating KPI targets is no greater than a network-specified probability  $\alpha$. More generally, one may wish to guarantee constraints of the form $\mathbb{E}[f(\textrm{KPI})]\leq\alpha$, where $f(\cdot)$ is a function of the KPIs.

 \noindent $\bullet$ \emph{Deployment-time calibration}: The statistical guarantees provided by pre-deployment calibration may be practically violated at run-time owing to \emph{statistical shifts} between nominal and actual conditions, or  due to \emph{conflicts} between AI apps that are deployed at the same time  despite having been calibrated separately  \cite{delprever2024pacifista}. Via deployment-time calibration, the controller  continuously monitors the KPIs, adjusting the post-processing operation and/or the hyperparameters in response to shifting data distributions or unforeseen situations. Given a maximum tolerated target level $\alpha$, the goal is  to ensure  KPI levels over time, even as conditions change, attaining \emph{deterministic} guarantees of the form $\sum_{t=1}^{T} f(\textrm{KPI at time } t )/T \leq \alpha + o(1)$, where $o(1)$ tends to zero as $T\rightarrow \infty$ and time $t$ runs over a relevant time scale.
 
 \noindent $\bullet$ \emph{Post-deployment counterfactual analysis}: After deploying an app and measuring the relevant KPIs, the controller may be interested in assessing the potential performance that \emph{would have} been obtained with different hyperparameters or AI apps. This type of \emph{counterfactual analysis} -- evaluating potential outcomes that were in fact not  realized -- is essential to diagnose possible performance and improve efficiency.

\subsection{Overview}
This paper reviews recent advances in pre-deployment calibration, deployment-time calibration, and post-deployment counterfactual analysis by outlining the underlying statistical frameworks and by presenting several exemplifying applications -- from channel prediction to scheduling. 

The main thread that runs through this line of work is methodological: formal calibration methods rely on modern statistical tools grounded in \emph{conformal prediction} and recent extensions thereof 
\cite{angelopoulos2024theoretical}.  Conformal prediction is a statistical framework introduced in the late 90s  for  \emph{post-hoc calibration}. Post-hoc calibration aims at ensuring  {formal reliability guarantees} for black-box models via the post-processing of a model's outputs. Recent extensions  have broadened the applicability of conformal prediction tools to the selection of hyperparameters \cite{LTT,ramdas2024hypothesis}.  



The next sections cover each one aspect of calibration: pre-deployment calibration, deployment-time calibration, and post-deployment counterfactual analysis. Depending on the given setting, pre-deployment calibration can be carried out either via uncertainty quantification or via hyperparameter selection, and the next two sections cover separately these two approaches.

\section{Pre-Deployment Calibration via Uncertainty Quantification
}

AI applications frequently function as \emph{predictors}, producing outputs that guide downstream components in making decisions and optimizing system configurations. For instance, an AI application may forecast packet arrivals for ultra-reliable low-latency communication (URLLC) traffic, enabling proactive resource allocation. Or an AI model may predict channel conditions to support efficient the allocation of communication resources such as beams or time-frequency slots \cite{zecchin2024forking}. This section focuses on the challenge of pre-deployment calibration in such \emph{predict-then-act} systems.


\subsection{Prediction-Based Decision Making}

Consider the problem of optimizing system settings $x$, such as scheduling decisions or power levels, subject to constraints on KPIs. The KPI of interest, denoted by $R(x, y)$, generally depends not only on the decision variables $x$ but also on \emph{unknown} environmental or system variables $y$. Throughout this paper, we assume that all KPI measures -- represented by $R$ -- are negatively oriented, meaning that smaller values are preferable.

As an illustrative example, suppose $x$ represents a scheduling decision. One may wish to enforce a constraint on the latency  $R(x, y)$. However, the latency also depends on the interference level $y$, which is not known in advance by the scheduler.

We focus on scenarios where one aims to guarantee statistical KPI constraints of the form $\mathbb{E}_{y}[R(x, y)] \leq \alpha$, where the expectation $\mathbb{E}_{y}[\cdot]$ is taken over the random process $y$. In general, the distribution of $y$ is not known. Nevertheless, the system is equipped with a pre-trained AI-based predictor that provides estimates of the unknown variables $y$.

A conventional \emph{best-effort} approach leverages the AI predictor to generate a \emph{single-point estimate} $\hat{y}$ -- for example, a predicted interference level -- which is then fed into the optimization process. The system settings $x$ are selected to satisfy a deterministic constraint of the form $R(x, \hat{y}) \leq \alpha$. However, this approach offers no guarantee that the statistical KPI requirement $\mathbb{E}_{y}[R(x, y)] \leq \alpha$ is actually met by the chosen $x$.

\subsection{Reliable Prediction-Based Decision Making via Set Prediction}

As discussed next, the average KPI constraint  $\mathbb{E}_{y}[R(x, y)] \leq \alpha$ can be   met, even without making any assumption on the unknown process $y$, by moving beyond individual predictions towards \emph{post-processing} and \emph{set prediction}. To explain,  suppose that the controller can post-process the AI predictions to produce \emph{valid} ``error bars'' $\Gamma$. Accordingly, the post-processing output $\Gamma$  amounts to a subset  of possible values of the unknown process $y$ with the following property: the true value of the process $y$ is included in the subset $\Gamma$ with a controller-defined probability. Denoting the target \emph{coverage} probability as  $1-\beta$, this condition can be formally expressed as the inequality $\Pr[y\in\Gamma]\geq 1-\beta$. 

Depending on the task, the prediction set $\Gamma$ may take various forms: a subset of labels for classification; an interval -- or more generally, a union of connected regions -- for regression; a set of plausible future trajectories in time-series forecasting; or a collection of possible answers in the context of language models.

Given access to a prediction set $\Gamma$ with guaranteed coverage probability $1 - \beta$, the optimizer can now replace the intractable average constraint $\mathbb{E}_{y}[R(x, y)] \leq \alpha$ with the known \emph{worst-case} constraint $\max_{y\in\Gamma}R(\ensuremath{x,y})\leq\gamma$. 
By construction, this guarantees that the KPI $R(x, y)$ remains below the threshold $\gamma$ with probability at least $1 - \beta$. Through an  appropriate selection of the parameters $\beta$ and $\gamma$, it can be shown that the original average constraint is satisfied \cite{zecchin2024forking}. Indeed, recent results demonstrate that this approach is \emph{optimal} for a broad class of optimization problems with statistical constraints under unknown distributions \cite{kiyani2025}.

\subsection{Reliable Set Prediction via Conformal Prediction}

\begin{figure}
    \centering
    \includegraphics[width=0.65\textwidth]{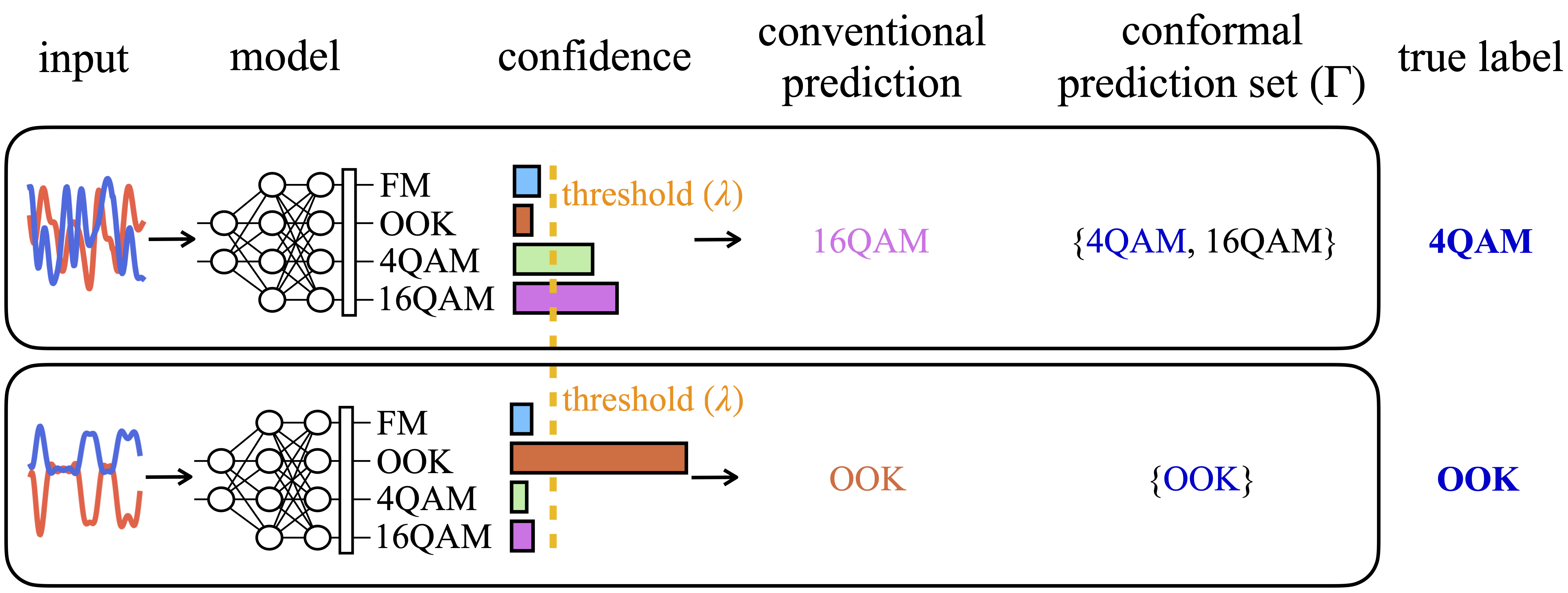}
    \caption{ Conformal prediction is a post-processing method that  produces decision sets $\Gamma$ that include all output values $y$ to which the AI app assigns  a confidence level above a given threshold $\lambda$. By  optimizing the threshold $\lambda$ on the confidence levels of the AI app, conformal prediction  guarantees  that the prediction set $\Gamma$ covers  the true output with a user-defined probability.} 
    \label{fig:CP}
\end{figure}

Conformal prediction is a post-processing technique that transforms the output of any black-box predictor into a prediction set $\Gamma$ with user-defined coverage guarantees. As illustrated in Figure~\ref{fig:CP}, conformal prediction constructs the set $\Gamma$ by including all candidate values $y$ to which the AI model assigns a confidence score above a given threshold $\lambda$. This adaptive approach ensures that the size of the prediction set $\Gamma$ reflects the difficulty of the current input $x$: ``easy'' inputs yield smaller sets, while ``difficult'' inputs result in larger, more cautious predictions.


In an ideal scenario, the model's confidence scores would align perfectly with the ground-truth data distribution. Under such conditions, the threshold $\lambda$ could be directly calibrated based on the model’s internal confidence levels. However, achieving this alignment requires strong distributional assumptions, which are rarely met in practice \cite{angelopoulos2024theoretical} -- a limitation that may be interpreted as a manifestation of the \emph{no-free-lunch} theorem.

To overcome this challenge, conformal prediction uses a held-out calibration dataset to empirically determine the threshold $\lambda$. The key idea is to estimate the distribution of model errors on the calibration set and set $\lambda$ accordingly, ensuring that the resulting prediction set $\Gamma$ satisfies the coverage condition $\Pr[y \in \Gamma]\geq 1-\beta$ 
for any user-specified confidence level $1 - \beta$, regardless of the underlying model’s accuracy. This guarantee holds under a single assumption: that the calibration and test data are \emph{exchangeable}. For instance, they can be independent and identically distributed (i.i.d). Exchangeability thus defines the notion of \emph{nominal} test conditions: conditions under which test inputs are drawn from the same statistical population as the calibration data.

Importantly, while conformal prediction ensures validity (i.e., guaranteed coverage), it does not guarantee \emph{informativeness}. If the underlying model is inaccurate, the prediction set $\Gamma$ may become excessively large in order to maintain the desired coverage level \cite{zecchin2024generalization}. In decision-making applications, such conservative prediction sets can make the worst-case constraint $\max_{y \in \Gamma}R(x, y)\leq\gamma$ difficult to satisfy without over-allocating system resources to hedge against uncertainty.

It is therefore desirable to tighten prediction sets as much as possible without violating coverage guarantees. One promising direction is to improve the quality of the confidence scores extracted from the AI model’s output, without modifying the model itself. The following example illustrates this approach.

 \subsection{An Application: Power Control for Unlicensed Spectrum Access}

Consider a spectrum access setting  in which an unlicensed  device must allocate transmission power  while respecting a constraint on the interference towards an incumbent, licensed, receiver. The interference level caused to the licensed receiver   depends on the channel from the unlicensed transmitter  to the licensed receiver, but this channel  is unknown to the unlicensed user. Thus, the unlicensed device relies on an AI-based prediction of the future realizations of the unlicensed-to-licensed  channel gain $y$ to set its transmission power $x$. The goal is to satisfy an interference constraint of the form $\mathbb{E}_{y}[R(\ensuremath{x,y})]\leq\alpha$, where 
$R(x,y)$ is the instantaneous interference level caused to the licensed receiver.

Using the approach explained in this section, the unlicensed user can leverage conformal prediction to ensure that the interference constraint is provably satisfied. It is, however, critically important for the prediction set produced by conformal prediction not to be too large, since the size of the set determines the degree to which the transmission power of the unlicensed user must be reduced in order to avoid excessive inference.

 \begin{figure}
    \centering
    \includegraphics[width=0.55\textwidth]{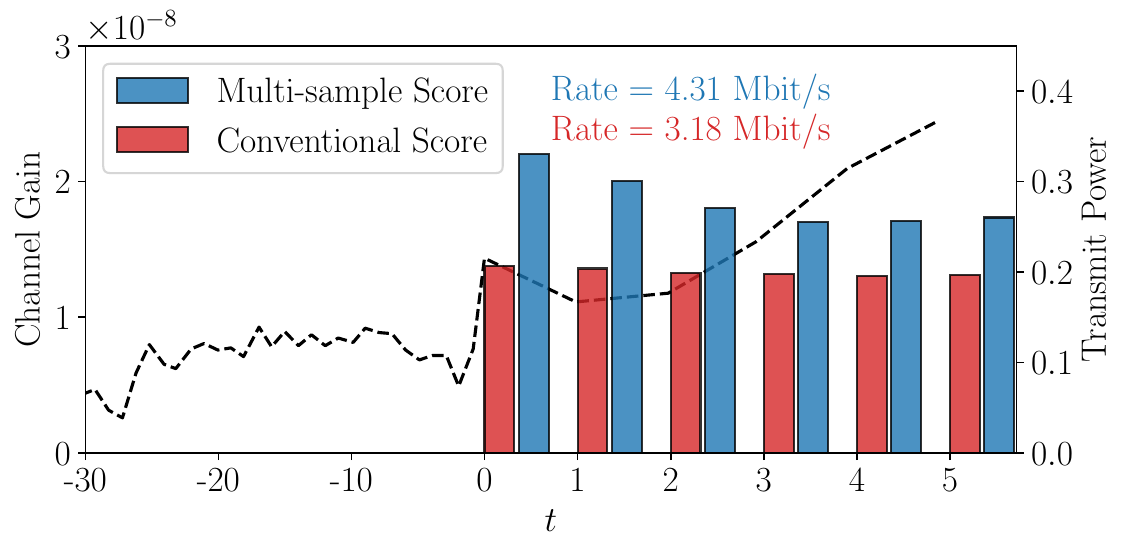}
    \caption{True channel gain of the licensed user (dashed line), which is unknown and must be predicted after time zero, and (normalized) powers allocated by leveraging conformal prediction with two different confidence scores: a standard (negative) squared loss    and a multi-sample score introduced in \cite{zecchin2024forking} that yields multi-modal sets. Although both scores are based on the same AI model, a score that is better tailored to the problem at hand can yield significantly less conservative solutions (i.e., larger transmission powers).}
    \label{fig:ExamplePC}
\end{figure}

Figure \ref{fig:ExamplePC} depicts a realization of the channel towards the primary receiver, along with the unlicensed transmission powers obtained with the conformal prediction-based procedure presented in this section. Specifically, we fix the AI predictor and evaluate the confidence levels in two different ways. The first, conventional, approach uses a single prediction $\hat{y}$, and evaluates the confidence assigned by the model to any possible channel trajectory $y$ as the standard negative squared loss $-\lVert y-\hat{y}\rVert^2$. Upon thresholding, this score yields a single contiguous prediction interval (e.g., one range of values for the next channel gain). As seen in the figure, this method yields  an overly conservative power allocation (low transmit power) to cover the worst-case channel outcome,  sacrificing the unlicensed user’s rate. 

The unimodal interval obtained via the standard squared loss score can indeed be too broad if the future channel gains have a \emph{multi-modal} distribution. This is often likely to be the case, since, e.g., the channel could either remain stable or drop sharply, depending on random connectivity conditions. The second approach depicted in Figure \ref{fig:ExamplePC} leverages the probabilistic nature of the AI predictor (which may be an autoregressive model like a transformer) to  sample several plausible future channel trajectories, $\{\hat{y}_m\}_{m=1}^M$. Then, confidence in a possible channel evolution $y$ is measured as $-\min_{m\in\{1,...,M\}}\lVert y-\hat{y}_{m}\rVert^2$. This yields {multi-modal} prediction  intervals that are significantly smaller in total size than a single large interval covering all modes. As illustrated in Figure \ref{fig:ExamplePC}, this yields markedly less conservative power allocations that support a higher transmission rate for the unlicensed user.

Overall, this example illustrates how, even given the same AI model, extracting confidence scores in ways that are better informed about the nature of the problem at hand can vastly improve efficiency, while maintaining reliability.

\section{Pre-Deployment Calibration via Hyperparameter Selection}

The pre-deployment methodology reviewed in the previous section applies to any \emph{predict-then-optimize} problem in which an AI predictor is used to inform the optimization of communication settings. In more general scenarios, statistical KPI guarantees can be guaranteed under nominal conditions   via \emph{hyperparameter selection} prior to deployment.

As discussed, hyperparameters consist of settings that are exposed to the controller by the API  of a pre-trained AI app. Conventionally, hyperparameters are chosen through techniques like grid search or Bayesian optimization, which aim to maximize validation-set accuracy or minimize validation-set loss. Such approaches are of \emph{best-effort} nature: they target average performance via empirical measures, but do not provide formal guarantees that the selected model will meet any specific statistical reliability threshold. This section reviews pre-deployment calibration strategies based on hyperparameter selection that offers statistical KPI guarantees under nominal conditions.

\subsection{Reliable Hyperparameter Selection via Multiple Hypothesis Testing}

The \emph{learn-then-test} (LTT) framework is a recently introduced statistical methodology that formulates hyperparameter selection as a multiple hypothesis testing problem, enabling the identification of hyperparameter configurations with provable risk control \cite{LTT}.

To illustrate the intuition behind LTT, consider an apparently unrelated scenario: an online shopping platform aiming to optimize its website to enhance user engagement. The company evaluates multiple design features -- such as color schemes, layout styles, and logos -- by simultaneously testing a set of hypotheses. For each feature, the \emph{null hypothesis} posits that the current design is preferable, while the alternative hypothesis suggests that a different configuration may yield better customer engagement.

Based on data collected from user interactions, the company must decide, for each feature, whether to retain the current design (i.e., accept the null hypothesis) or adopt a new one (i.e., reject the null). Rejections of the null hypothesis, indicating a preference for a new configuration, are referred to as \emph{discoveries}.

Standard statistical procedures ensure control over the \emph{family-wise error rate} (FWER) -- that is, they bound the probability of making even a single false discovery (an incorrect rejection of a true null hypothesis) by a user-specified threshold $\beta$ as $
\Pr[\text{false discovery]}\leq\beta$.

Returning to the problem of hyperparameter selection, let $\lambda$ denote a candidate hyperparameter configuration -- for example, a specific setting of fairness parameters in a scheduling application. We aim to enforce a reliability constraint on a negatively oriented KPI, or loss function, denoted $R(\lambda)$. Specifically, we require that the probability of $R(\lambda)$ exceeding a maximum tolerable threshold $\alpha$ remains below a target level $\beta$: $ \Pr[R(\lambda)>\alpha]\leq\beta$. As an example, one may wish to ensure that the system latency -- modeled by $R(\lambda)$ -- exceeds $\alpha = 1$ ms with probability no greater than $\beta = 10^{-5}$.

The core idea behind the LTT framework is to associate each candidate hyperparameter $\lambda$ with the null hypothesis that it is unreliable -- that is, that the condition $R(\lambda) > \alpha$ holds. Rejecting this null hypothesis is thus interpreted as a discovery of a reliable hyperparameter configuration.

By applying a multiple hypothesis testing procedure that controls the FWER, LTT guarantees that the probability of returning any unreliable hyperparameter is bounded by the user-defined level $\beta$, thereby satisfying the desired statistical guarantee $\Pr[R(\lambda) > \alpha] \leq \beta$.

In practice, the testing process involves collecting observations of the KPI values $R(\lambda)$ for each candidate configuration and computing a test statistic that quantifies the strength of evidence against the null hypothesis of unreliability. These test statistics may be based on traditional \emph{p-values}, or on more robust and flexible \emph{e-values}, which will be discussed in further detail below \cite{ramdas2024hypothesis}. The evaluation of such test variables typically involves simple statistical operations such as computing empirical averages and applying non-linear transformations -- as in the case of the Hoeﬀding-Bentkus p-value \cite{LTT}.


\subsection{Adaptive Hyperparameter Selection via Multiple Hypothesis Testing}

LTT is that it is a \emph{batch} procedure: it requires evaluating KPIs for all candidate hyperparameters at once. In practice, the space of hyperparameters can be large, and each KPI evaluation may be expensive. For instance, evaluating the latency attained by some hyperparameter settings for scheduling may require large network simulations and/or interactions with the real world for a field trial. To improve efficiency, it would be useful to carry out testing sequentially, discarding less performing hyperparameters early on and terminating the hyperparameter selection process as soon as possible.

An adaptive version of LTT with these features was introduced in \cite{zecchin2024adaptive} under the name \emph{adaptive LTT (aLTT)}. aLTT conducts the testing in sequential rounds rather than all at once, supporting optional continuation and adaptive termination.

At each evaluation round, aLTT evaluates the instantaneous values of the KPIs $R(\lambda)$ for a subset of the candidate hyperparameters. Candidates that appear unreliable can be dropped early, focusing subsequent rounds on the more promising ones. The process continues until it either identifies a sufficiently large set of reliable hyperparameters or it runs out of time.  Crucially, aLTT is designed to maintain the same statistical guarantees  as the original batch LTT. 

To this end, aLTT leverages test variables built from \emph{e-values}. An e-value is a generalization of a likelihood ratio that bears the interpretation of \emph{wealth growth} in the \emph{testing-by-betting} framework. To explain, imagine trying to test the
null hypothesis that a roulette at a casino is \emph{fair}. If you can gain
money, on average, by playing the roulette, then you have proved that the roulette  is
not fair. What is more, the amount of money you have gained provides a direct, quantitative, measure of the evidence against the null hypothesis that the roulette is fair. Accordingly, an e-value -- measuring the wealth growth -- provides evidence on the reliability of a hyperparameter configuration. Being determined by betting strategies, e-values  may be optimized online using tools such as online gradient descent and variants \cite{ramdas2024hypothesis}. 

\subsection{An Application: Hyperparameter Selection for Downlink Scheduling }

\begin{figure}
    \centering
    \includegraphics[width=0.5\textwidth]{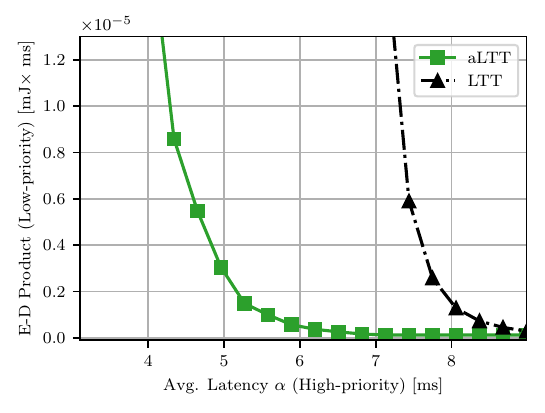}
    \caption{Energy-delay (E-D) product for low-priority UEs versus the target average latency for high-priority UEs attained by LTT and aLTT. Both pre-deployment calibration methods provably meet the average latency requirements. However,  aLTT can use calibration data more efficiently to identify better-performing solutions for the low-priority UEs.}
    \label{fig:aLTT}
\end{figure}

Consider an AI-driven downlink scheduler that has two key hyperparameters in the configuration vector $\lambda$: one determining the fairness across user equipments (UEs)  and another controlling the transmission power level. The controller wishes to ensure that the average latency $R(\lambda)$ of high-priority UEs exceeds some level $\alpha$ (ms) with a probability no larger than $\beta=0.1$.  Among the configurations that satisfy this requirement, the controller wishes to minimize the energy-delay product -- an overall measure of transmission efficiency -- for lower-priority UEs. This optimization is done in a best-effort manner using calibration data. 

Figure \ref{fig:aLTT} compares the energy-delay product for the lower-priority UEs obtained by LTT and aLTT as a function of the latency constraint for the high-priority UEs. Both schemes are compared for the same amount of calibration data. While both aLTT and LTT provably satisfy the latency reliability requirements for high-priority UEs, the conventional, non-adaptive, LTT strategy yields much higher energy-delay products. In fact, levels of latency lower than 7 ms are not even attainable by LTT. 

Overall, this example shows the importance of deploying powerful pre-deployment calibration strategies in order to use the available AI models and data in the most efficient way, enhancing auxiliary KPIs while provably guaranteeing critical KPIs. 
 
\section{Deployment-Time Calibration}
Even after a model is calibrated during a pre-deployment phase to ensure KPI performance under nominal conditions, the system behavior can deviate from the conditions assumed during calibration due to statistical shifts and/or inter-app conflicts, causing a violation of KPI constraints. This section introduces a class of calibration methods that operate online, at deployment time, with the goal of ensuring worst-case, deterministic, KPI guarantees.

\subsection{Robust and Reliable Prediction-Based Decision Making}

To illustrate the operation of deployment-time calibration, consider the setting studied in Section II in which an AI-based predictor is used to inform a decision-making block, e.g., for resource allocation. Using pre-deployment calibration, one can ensure statistical guarantees of the form $\mathbb{E}[R]\leq \alpha$, where $R$ is the KPI of interest. However, these guarantees require calibration and test data to be exchangeable -- a condition that is violated when deployment conditions are different from the conditions under which calibration data were collected. Therefore, while pre-deployment calibration is still critical in facilitating the reliable deployment of AI models in wireless systems, online monitoring and updating mechanisms are also necessary.

Deployment-time calibration aims at guaranteeing deterministic constraints of the form $\sum_{t=1}^{T} R_t/T \leq \alpha + o(1)$, where $R_t$ is the relevant KPI measured at time $t$. The discrete time $t$ ranges over the time scale of interest. For example, for apps run by a near-real time controller, $t$ may run over transmission time intervals (TTIs). This condition ensures a time-averaged performance that deviates by a vanishingly small amount from the target level $\alpha$. Importantly, this condition must hold even when the sequence of data is adversarially selected, thus allowing for any type of non-stationary behavior.

Recent work in statistics and optimization has introduced a class of techniques that can offer this type of guarantees  by leveraging feedback about past KPI levels $R_t'$ with $t'\leq t$ to update the threshold $\lambda_t$ used in determining the prediction set $\Gamma_t$ at time $t$ (see Section II-C) \cite{gibbs2021adaptive}. The approach, referred to as \emph{online conformal prediction}, is formally grounded in online convex optimization, and is conceptually and computationally straightforward.  

The core principle behind online conformal prediction is to treat each new KPI observation as an opportunity to recalibrate. After each prediction is made and the KPI $R_t$ is measured, the basic threshold update rule works as follows: $\lambda_{t+1} = \lambda_t - \eta_t (R_t-\alpha)$, where $\eta_t$ is a step size. Thanks to this update, if the KPI $R_t$ -- e.g., the latency level --  happens to be higher than the target $\alpha$ at time $t$, the threshold $\lambda_{t+1}$ is lowered. This ensures that  the prediction set will tend to increase in size (see Figure 2), yielding more conservative decisions at the next time step. Assuming that the KPI of interest is decreasing in the size of the prediction set, this reactive behavior can be shown to ensure the satisfaction of the mentioned time-averaged deterministic constraints.

Online conformal prediction applies the same threshold $\lambda_t$ to all inputs that may be processed by the prediction app at time $t$. In practice, in wireless systems,  one may wish to differentiate the level of conservativeness of the prediction set as a function of the input. A recent extension of online conformal prediction, referred to here as \emph{localized} online conformal prediction, enables finer control over the threshold $\lambda_t$, allowing it to be a function of the input $x$ via a function $\lambda_t(x)$. 

\subsection{An Application: Downlink Beam Selection}

Consider a frequency-division duplex (FDD) massive multi-antenna system in which downlink beams are selected from a predefined codebook based on the transmission of downlink pilot signals. The process begins with the base station selecting a candidate subset of beams on which to transmit pilot signals to the UEs. Upon receiving these pilots, the UEs estimate the signal-to-noise ratio (SNR) for each candidate beam and report this information back to the base station. Based on the feedback, the base station then selects the most suitable beams from within the candidate subset. The selection of the initial candidate subset is a critical step: if optimal beams are excluded at this stage, the base station is unable to identify and utilize them, potentially compromising the quality of service delivered to the UEs.

\begin{figure}
    \centering
    \includegraphics[width=0.5\textwidth]{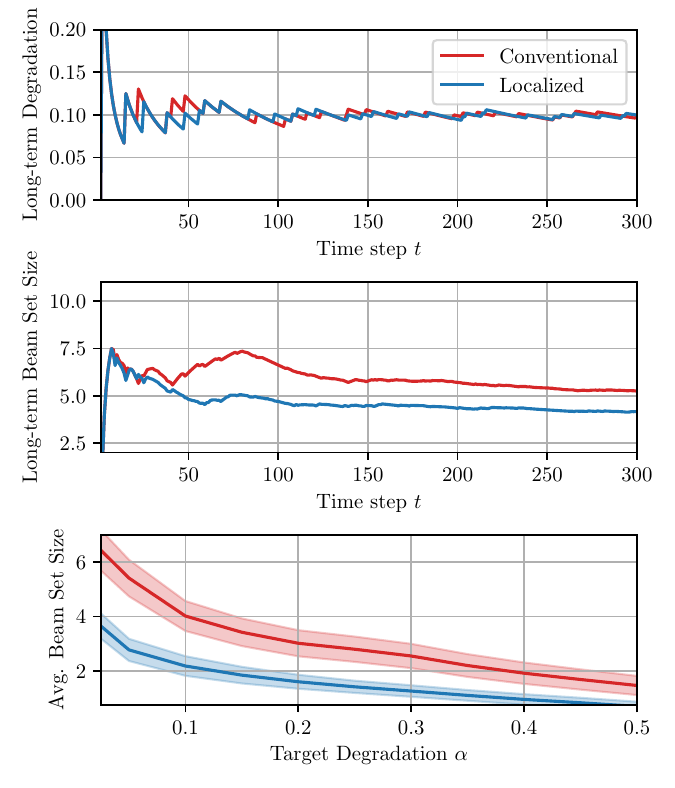}
    \caption{Cumulative normalized SNR degradation with respect to the optimal beam and corresponding beam set size as a function of time for a target normalized SNR degradation  $\alpha=0.1$ (top two figures); and average beam set size, evaluated after $T=10000$ steps, as a function of the target cumulative normalized SNR degradation (bottom figure). While both conventional and localized online conformal prediction meet the desired cumulative SNR degradation level as time goes on, the localized approach yields significantly smaller candidate beam set sizes, reducing training overhead.}
    \label{fig:ARC}
\end{figure}

A way to inform the selection of the candidate beams is via prediction. Using information from past observed SNR levels and contextual information such as the UE location, the base station can predict what the best beams are likely to be, selecting them as candidates for downlink training. Given an AI-based predictor of per-beam SNR levels, or a predictor of optimal beams, deployment-time calibration can maintain a threshold $\lambda_t$ that determines which beams are included in the candidate set. More conservative choices, i.e., smaller values of $\lambda_t$, yield larger sets.

 Assume as the reliability measure $R_t$ the normalized level of SNR degradation as compared to the optimal beam within the codebook. This quantity may be periodically evaluated via full-codebook training, and the intermittent availability of feedback can be addressed as explained in \cite{wang2025mirror}. For a degradation level $\alpha=0.1$, Figure \ref{fig:ARC} shows the cumulative average SNR degradation and the corresponding beam set size over time (top two figures) and the average candidate beam size as a function of the maximum tolerated SNR degradation (bottom figure) for online conformal prediction and localized online conformal prediction. As highlighted in the plot, both schemes can attain the desired time-averaged value of SNR degradation, with localized online conformal prediction yielding smaller candidate set sizes. This in turn entails the need to transmit fewer pilots. 

Overall, this example confirms the effectiveness of deployment-time calibration methods in ensuring KPI guarantees irrespective of statistical shifts between calibration and deployment conditions.

\section{Post-Deployment Counterfactual Analysis}

After an AI-driven app is deployed in a network and KPI values are observed over a period of time, the controller may wish to run some diagnostic tests to assess the degree to which the app is optimized for the given deployment conditions.  In this regard, a natural question to address is: ``\emph{What if} we had chosen a different app?''  For example, the controller may ask ``If we had used a different scheduling algorithm during last hour’s traffic surge, would the queueing backlog have been lower?'' Addressing such \emph{counterfactual} questions is crucial for retrospective performance evaluation and for optimizing app deployment policies. 

\subsection{Reliable Counterfactual Analysis via Set Prediction}

Answering counterfactual queries is inherently challenging due to selection bias: we only observe the KPIs for the apps that were actually selected, not for those that were not. Naturally, we cannot rewind time and re-run the same scenario using an alternative app under identical conditions. This is precisely where counterfactual analysis tools become essential.

One principled approach to counterfactual analysis involves conformal prediction  \cite{lei2021conformal}. This method constructs prediction intervals $\Gamma$ for the unobserved KPIs, ensuring that the true KPI lies within the interval with high probability $1-\beta$, i.e., $\Pr[\textrm{KPI}\in \Gamma]\geq 1-\beta$. The key innovation over standard conformal prediction (as discussed in Section II) lies in reweighting the calibration data for the alternative app of interest. This reweighting accounts for the traffic and connectivity conditions in the target scenario -- i.e., the scenario in which a different app was actually deployed. If certain conditions are under-represented in the data for the alternative app (because that app was rarely used in those conditions), they are assigned greater weight when computing the conformity threshold $\lambda$ \cite{hou2025if}.

\subsection{An Application: Counterfactual Analysis of Scheduling Apps}

Consider a downlink scenario in which the base station can choose between two scheduling strategies: \emph{round robin} (RR), which ensures fairness but does not account for channel conditions, and \emph{proportional fair channel-aware}  (PFCA) scheduling, which leverages channel state information to optimize throughput but may result in uneven performance across UEs. The network controller selects between RR and PFCA based on a policy informed by the initial system state, such as queue backlogs and connectivity conditions. For instance, RR may be chosen if the base station predicts that it will yield uniformly small residual backlogs across UEs within a specified time horizon.

\begin{figure}
    \centering
    \includegraphics[width=0.5\textwidth]{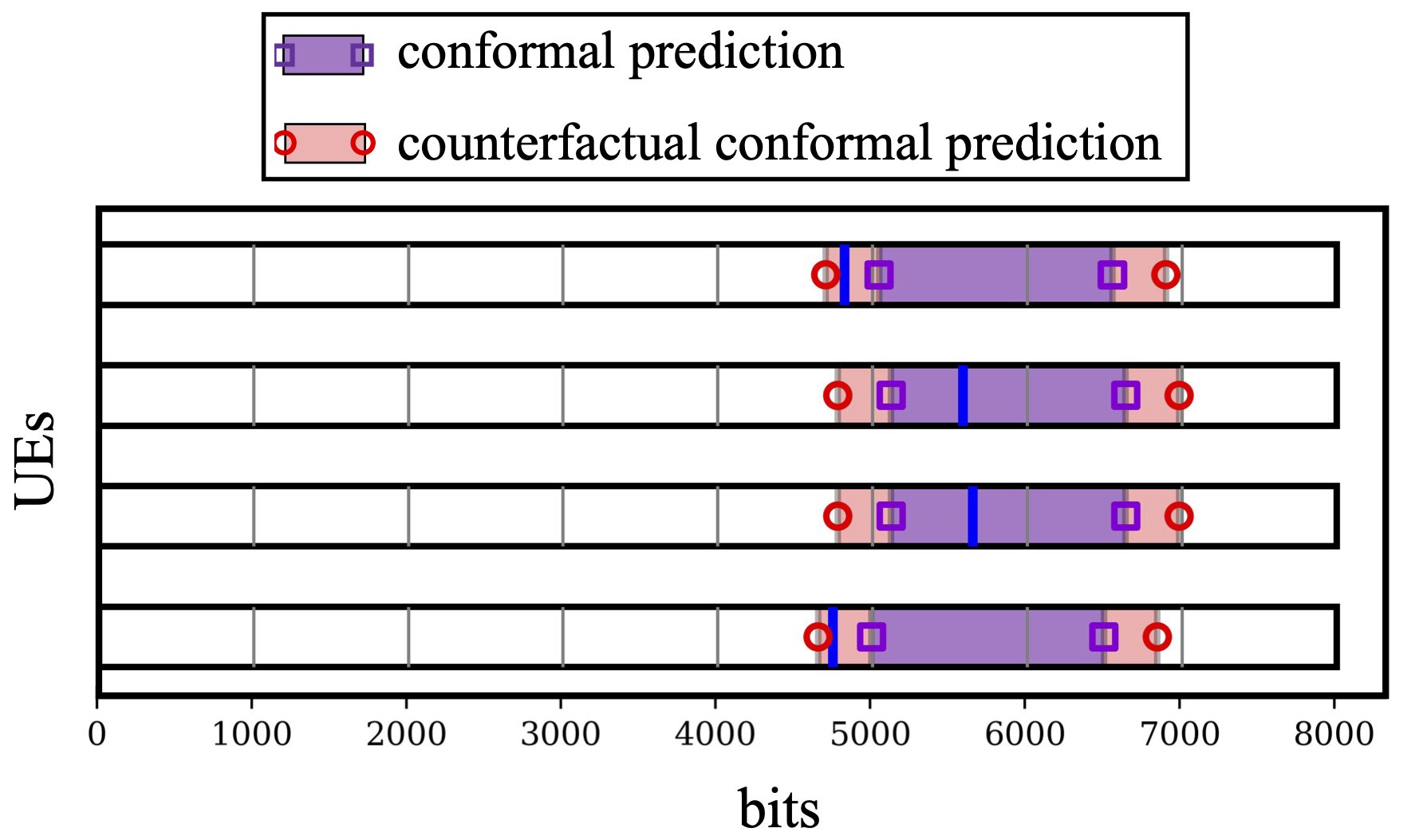}
    \caption{Illustration of prediction intervals for the final backlogs at the UEs obtained after the deployment of an alternative app using conventional conformal prediction and  counterfactual conformal prediction. Blue bars represent the true counterfactual backlogs. }
    \label{fig:CCKE}
\end{figure}

Figure \ref{fig:CCKE} illustrates the true final queue backlogs for the UEs, along with the prediction intervals generated by standard conformal prediction and counterfactual conformal prediction. Unlike the latter, standard conformal prediction does not adjust for the differing distributions of contexts under which each scheduling strategy is selected. As shown in the figure, this mismatch can lead standard conformal prediction to fail in covering the true final backlogs. In contrast, counterfactual conformal prediction -- by accounting for selection bias -- is statistically guaranteed (up to a user-specified probability) to provide valid coverage for the final queue backlogs.



\section{Conclusions}

Across the AI deployment lifecycle -- from initial development to live operation and retrospective analysis --  conformal calibration techniques provide a suite of tools to ensure reliability and trustworthiness of black-box models in wireless systems. We have surveyed how these methods can be applied for pre-deployment uncertainty quantification and hyperparameter selection, deployment-time adaptation, and post-deployment counterfactual performance evaluation. Conformal calibration is  model-agnostic -- it can wrap around any AI app, making it a flexible add-on for existing AI solutions. Through this approach, network operators would gain the confidence that even though an AI model may be a complex black box, its outputs are accompanied by guarantees: errors are bounded by design, and any deviation is caught and corrected in time. 


The calibration schemes presented here can be often improved by incorporating \emph{contextual} information. In fact, unlike generic machine learning applications, in wireless systems, calibration can leverage not only basic measurements --  typically reporting KPI levels -- but also information about the current context, including traffic and mobility levels, as well as connectivity conditions \cite{yoo2025calibrating}.

Ultimately, conformal calibration techniques can help transform AI from a potential reliability concern to a dependable enabling technology for next-generation wireless systems.


\bibliographystyle{IEEEtran}
\bibliography{biblio}

\end{document}